\documentclass[superscriptaddress,twocolumn]{revtex4-2}
\usepackage{graphicx}
\usepackage{dcolumn}
\usepackage{bm}
\usepackage{color}
\usepackage{ulem}
\usepackage{amsmath}
\usepackage{amsfonts}
\usepackage{amssymb}
\usepackage{tikz}
\usepackage{lineno}
\usepackage[]{changes}
\usepackage{txfonts}
\usepackage{comment}
\usepackage[colorlinks,linkcolor=blue,anchorcolor=blue,citecolor=blue,
bookmarksnumbered]{hyperref}

\definechangesauthor[name={2nd}, color=orange]{2nd}

\begin{document}

%\linenumbers

\title{Controllable quantum scars induced by spin-orbit couplings in quantum dots}
\author{Lin Zhang}
\author{Yutao Hu}
\author{Zhao Yao}
\author{Xiaochi Liu}\email{liuxiaochi@csu.edu.cn}
\author{Wenchen Luo}\email{luo.wenchen@csu.edu.cn}
\author{Kehui Sun}
\affiliation{School of Physics,
%\&Hunan Key Laboratory of Nanophotonics and Devices,
Central South University, Changsha 410083, China}
\author{Tapash Chakraborty}
\affiliation{Department of Physics and Astronomy, University of Manitoba, Winnipeg,
Canada R3T 2N2}
\date{\today }

\begin{abstract}
Spin-orbit couplings (SOCs), originating from the relativistic corrections  
in the Dirac equation, offer nonlinearity in the classical limit and are capable of driving 
chaotic dynamics. In a nanoscale quantum dot confined by a two-dimensional parabolic potential 
with SOCs, various quantum scar states emerge quasi-periodically in the eigenstates of the system, when the ratio 
of confinement energies in the two directions is nearly commensurable. The 
scars, displaying both quantum interference and classical trajectory features on the 
electron density, due to relativistic effects, serve as a bridge between the 
classical and quantum behaviors of the system. When the strengths of  
Rashba and Dresselhaus SOCs are identical, the chaos in the classical limit 
is eliminated as the classical Hamilton's equations become linear, leading to the 
disappearance of all quantum scar states. Importantly, the quantum scars induced by SOCs are 
robust against small perturbations of system parameters. 
With precise control achievable through external gating, 
the quantum scar induced by Rashba SOC is fully controllable and detectable.
\end{abstract}

\maketitle

%%%%%%%%%%%%%%%%%%%%%%%%

\section{Introduction}

Quantum scars which manifest as the localization behavior displaying certain unstable classical 
periodic orbits exist in the high-energy levels in the quantum system with chaotic 
dynamics being driven in its classical limit. The quantum scar was first discovered 
while studying the quantum eigenstates of the stadium billiard model which drives chaotic 
dynamics in the corresponding classical model \cite{quantumscar1} and later was named as such 
by Heller \cite{heller}. Quantum scarring has thus far drawn great attention and interest 
\cite{books} and has been observed experimentally in various systems, including quantum well 
and microwave resonators \cite{exp1}. The localization nature 
of quantum scarring without participation of the many-body system is convenient to be 
applied and attracts interest across various fields. On the other hand, the quantum many-body scars 
localizing eigenstates to prevent thermalization are expected to be useful in quantum 
computing \cite{qmbscar}. 

Recently, the perturbation induced quantum scars have been studied in quantum 
dot (QD) systems confined at the semiconductor heterostructure with or without 
an external magnetic field \cite{piscar1,piscar2,qls}. These quantum scars are 
induced by a bunch of impurities which make the (nearly) degenerate states of 
the QD resonant to localize the electron density along the underlying classical trajectories. 
As an artificial atom \cite{maksym}, the low-dimensional QD \cite{QD_book, QD01, 
QD03,QD04,QD05} offers an ideal platform for controlling both the spin and the charge of 
single or multiple electrons. The parabolic confinements of QDs render the system 
a two-dimensional (2D) quantum harmonic oscillator which holds practical and 
fundamental significance in physics. The quantum scars found in QDs also reveal 
profound connections between the classical and the quantum systems. 

Both nonrelativistic and relativistic quantum systems have been found to possess 
quantum scars \cite{relativeQS}. The focus has also been on quantum scarring in 
relativistic quantum systems which are described by the Dirac equation, especially in 
graphene systems \cite{relativeQS_graphene}. However, the experiments 
in monolayer and bilayer graphene to explore quantum chaos have not been successful \cite{graphene_exp}. On the other hand,
the spin-orbit coupling (SOC) is also a 
relativistic effect originating from the Dirac equation. Its 
corresponding classical Hamiltonian leads to nonlinearity in Hamilton's 
equation and it is possible to drive chaotic dynamics \cite{SOCchaos}. 
Exploring quantum scars induced by SOCs could thus offer an intriguing avenue
\cite{Berger}. 

The studies on QDs with Rashba SOC or/and Dresselhaus SOC have been 
reported extensively thus far \cite{QDSOC01,QDSOC02,QDSOC03,QDSOC04,
QDSOC05,QDSOC06,QDSOC07,QDSOC08,QDSOC09,QDSOC10,QDSOC12,
QDSOC13,QDSOC14,QDSOC15,QDSOC16,Intronati}. 
The ground states of QDs with SOCs have been studied to explore 
topological nontrivial features in spin fields \cite{luo1,luo2,skyrmionQD,luo3}. 
Vortex-like spin textures in the 
ground states carry different topological charges induced by Rashba SOC or 
linear Dresselhaus SOC. Considering that the Rashba SOC can be conveniently 
tuned via an external gate \cite{Rash01,Rash02,Rash03,Rash04,Rash05}, 
the spin textured ground states could have potential applications in 
spintronics and quantum information \cite{Zutic,Smejkal,Sinova}. Yet, the 
excited states in QDs with SOCs have not been sufficiently studied, especially in 
the energy region containing classical chaos. 

Here we investigate the excited states as well as the quantum scarring in 
spin-orbit coupled QDs. The scars can appear in the eigenstates quasi-periodically 
(the period is not fixed and gradually increased with the eigenenergy). 
We also confirm that the condition of scarring in the quantum states exactly follows the 
chaos condition in the classical limit. When the strengths of the Rashba and the 
Dresselhaus SOCs are equal, the classical Hamilton's equations are linear and no 
longer lead to chaos, hence there is no scar in the quantum system. Otherwise we observe 
various quantum scars depending on the systematic parameters. It is worth 
mentioning that the quantum scars induced by SOCs in QDs are highly robust against 
with small perturbations, unlike the classical chaotic behavior and could 
be referred to its quantum feature that the energies are discrete. Comparing with the 
impurities induced quantum scars, the scars induced by SOCs are more tunable, less 
random, exist at low-energy levels, and spin-involved. We thus expect the 
corresponding measurements to be more convenient by scanning 
tunneling spectroscopy \cite{stm}, scanning gate microscopy \cite{sgm}, scanning the NMR experiment \cite{sean} and the 
spin-dependent transport \cite{Berger,shenglin}.

\section{Model and formula}

The Hamiltonian of the quantum dot with both the Rashba and Dresselhaus SOCs 
is given by 
\begin{eqnarray}
&&\mathcal{H}=\frac{\mathbf{P}^{2}}{2m^{\ast }}+\frac{m^{\ast }}{2}\left(
\omega_x^2x^2+\omega_y^2y^2\right) +\frac{\Delta }2\sigma_z^{}+\mathcal{H}^{}_{SOC}, 
\label{qdh} \\
&&\mathcal{H}^{}_{SOC}=g_1^{}\left(\sigma_x^{}P_y^{}-\sigma_y^{}P_x^{}\right)+g_2^{}\left(
\sigma_y^{}P_y^{}-\sigma_x^{}P_x^{}\right),
\end{eqnarray}%
where $\omega^{}_x$ and $\omega^{}_y$ describe the parabolic confinements in the 
$x$ and $y$ dimensions, respectively. $\sigma^{}_i$ is the Pauli matrix and
the strengths of the Rashba and Dresselhaus SOCs are $g_1^{}$ and
$g_2^{}$ respectively. $P_i^{}=p_i^{}+eA_i^{}$ is the kinetic
momentum, where $e$ is the charge of an electron and the vector potential can be chosen in the symmetric gauge
$\mathbf{A}=\frac12 B\left(-y,x,0\right)$ with the magnetic field $B.$
The Zeeman term, which is the first order correction of the relativistic effect, is 
$\Delta =g\mu_B^{}B$, where $g$ is the Land\'{e} factor and $\mu_B$ is the Bohr 
magneton.

In \replaced{an}{the} adiabatic model the SOCs could have a classical correspondence \cite{SOCchaos} 
in the absence of a magnetic field,
\begin{eqnarray}
&&\mathcal{H}_{SOC} = (g_1 p_y - g_2 p_x) \sigma_x + (g_2 p_y - g_1 p_x) \sigma_y 
\notag \\ 
&&\rightarrow H^C_{SOC}= - \sqrt{(g_1 p_y - g_2 p_x) ^2 + (g_2 p_y - g_1 p_x) ^2},
\end{eqnarray}
which provides nonlinearity and is able to drive chaotic dynamics in the system. 
The full classical Hamiltonian reads
\begin{eqnarray}
H^C &=& \frac{p_x^2 + p_y^2}{2m^*}+ \frac{m^*}{2} (\omega_x^2 x^2 + \omega_y^2 y^2) 
\notag \\
 &-& \sqrt{g_1^2+g_2^2} \sqrt{p_x^2+p_y^2 -\frac{4g_1 g_2 }{g_1^2+g_2^2}p_x p_y }.
\end{eqnarray}
By solving the canonical equations, chaotic dynamics can appear when the SOC is anisotropic. 
If there is only Rashba SOC present, then $H^C_{SOC}=-g_1\sqrt{ p_x^2 + p_y^2}$. 
On the other hand, if only Dresselhaus SOC is present, then the classical 
correspondence is the same as that of Rashba SOC. It implies that whichever  
SOC is present, the classical behavior remains the same. Note that if the confinement 
trap is isotropic, classical trajectories in the phase space would be regular. The 
way leading to chaotic dynamics is to make the confinement anisotropic, which 
effectively makes the SOC anisotropic in the classical limit. Once chaos 
appears, the corresponding quantum scar induced by the SOC should be 
observed in the quantum dot. Considering the classical correspondence of the 
two types of SOCs being the same, the quantum scar would also be identical. 

The system is highly tunable, as both the Rashba SOC and confinements 
can be tuned by external gates, and the ratio of 
the Rashba SOC to the Dresselhaus SOC can be modified by 
applying an in-plane magnetic field \cite{Rash05}.
It is worth mentioning a special case where $g_1= \pm g_2$, i.e. the two SOCs 
are present simultaneously with equal strength. The classical correspondence 
becomes $H^C_{SOC}=-g_1 (p_x - p_y)$, which is a linear term in the 
Hamiltonian and does not lead to chaos. 

To study the quantum scar of the quantum dot system described by the 
Hamiltonian in Eq. (\ref{qdh}), the eigenstates are calculated in the exact 
diagonalization scheme. The Hamiltonian matrix is constructed in the basis 
of the two dimensional (2D) quantum oscillator whose Hamiltonian is 
$H_{0}=\frac{\mathbf{p}^{2}}{ 2m^{\ast}}+\frac{m^{\ast }}{2}
\left( \Omega _{x}^{2}x^{2}+\Omega_{y}^{2}y^{2}\right) +\frac{\Delta }{2}\sigma _{z}$, 
where the renormalized frequency is defined as 
$\Omega_{x,y}= \sqrt{\omega_{x,y}^2+ \omega_c^2/4}$ with the cyclotron 
frequency in a magnetic field being given by $\omega_c = |e|B/m^*$.
The basis of the 2D quantum oscillator is $|n \rangle \equiv |n_x,n_y, n_s \rangle$ where 
$n$ is a collective index marking the number of the basis, 
$n_s$ is the spin index in $n$, and $n_x$ and $n_y$ denote the two quantum numbers 
in two directions of the 2D quantum oscillator, respectively. The associated wave function 
of this basis is  
\begin{equation}
\psi _{n_{x},n_{y}}\left( \mathbf{r}\right)  =\frac{\exp \left( -\frac{%
x^{2}}{2\ell _{x}^{2}}-\frac{y^{2}}{2\ell _{y}^{2}}\right) }{\sqrt{%
2^{n_{x}+n_{y}}n_{x}!n_{y}!\pi \ell _{x}\ell _{y}}}H_{n_{x}}\left( \frac{x}{%
\ell _{x}}\right) H_{n_{y}}\left( \frac{y}{\ell _{y}}\right), 
\label{2Dbasis}
\end{equation}
where the natural lengths in the two directions are 
$
\ell _{x,y} =\sqrt{\hbar /m^*\Omega _{x,y}}.
$
In principle, there is no upper limit of $n_{x,y}$, so that the matrix of the Hamiltonian 
is infinity large. Practically, a truncation of $n_{x,y}$ is necessary and 
the low-energy states can be found accurately. 

Once the Hamiltonian (\ref{qdh}) is diagonalized, the $m-$th eigenstate can be 
expressed by the basis, $|\Psi_m \rangle= \sum_n C_n^m |n \rangle$, and its 
wave function is $\Psi_m (\mathbf{r}) = \langle \mathbf{r}| \Psi_m \rangle = \sum_n 
C^m_n \psi_{n_x, n_y}(\mathbf{r}) | n_s\rangle$. Here, $|n_s \rangle$ represents an 
eigenstate of $\sigma_z$, and thus the wave function $\Psi_m (\mathbf{r})$ is a 
two-component spinor. Generally, any observable field is given by
\begin{equation}
\Lambda (\mathbf{r}) =\Psi_m^\dag (\mathbf{r}) 
\Lambda \Psi_m (\mathbf{r}),
\end{equation}
where $\Lambda$ is the corresponding operator including the density operator 
$n$ (unity matrix), spin operators $\sigma_\mu$ with $\mu=x,y,z$ and etc. The 
current fields which are related to the spin fields are defined by 
$j_{x}\left( \mathbf{r}\right)  =-\frac{e}{m^{\ast }} \text{Re} \left[ \Psi_m
^{\dag }\left( \mathbf{r}\right) P_{x}\Psi_m \left( \mathbf{r}\right) \right]
+eg_{1}\sigma _{y}\left( \mathbf{r}\right) +eg_{2}\sigma _{x}\left( \mathbf{r%
}\right)$ and $j_{y}\left( \mathbf{r}\right)  =-\frac{e}{m^{\ast }}\text{Re}\left[ \Psi
^{\dag }\left( \mathbf{r}\right) P_{y}\Psi \left( \mathbf{r}\right) \right]
-eg_{1}\sigma _{x}\left( \mathbf{r}\right) -eg_{2}\sigma _{y}\left( \mathbf{r%
}\right)$.

Without loss of generality, we consider here the InAs quantum 
dots with the material parameters: the effective mass of electron is 
$m^{\ast}=0.042m_e$ where $m_e$ is the mass of free electron and Land\'{e} factor 
$g=-14$. The size of the QD is not fixed here, but we could 
fix the confinement in $x$ direction and vary the other confinement. The confinement 
lengths are defined by $R_i= \sqrt{\hbar / m^* \omega_i}$ and $R_x$ is fixed to $30$ 
nm associated with the characterized confinement energy $\hbar \omega_x = 2$ meV.

\section{Results}

\subsection{Isotropic QD with a single SOC}

When only one SOC is present and the quantum dot is isotropic, 
$\omega_x=\omega_y=\omega$, the associated classical Hamiltonian is 
$
H^C= \frac{p_x^2 + p_y^2}{2m^*}+ \frac{m^* \omega^2}{2} ( x^2 + y^2) 
- g_{1(2)} \sqrt{p_x^2 + p_y^2}, 
$
which does not lead to chaos \cite{SOCchaos}. In the quantum regime, the 
densities and the spin fields of all eigenstates in the isotropic quantum dot are 
deformed by the SOC. The rotational symmetry does not only exist in the ground 
state, but also exists in all excited states of the single-particle system, due 
to the symmetry $[L_z \pm \sigma_z/2, \mathcal{H}]=0$, where $L_z$ is the 
$z$ component of the angular momentum \cite{luo1}. The topological feature of 
the spin fields is also retained in the excited states, i.e. nontrivial patterns 
with nonzero topological charges are textured by the SOC.  

\begin{figure}[htbp]
\includegraphics[width=\columnwidth]{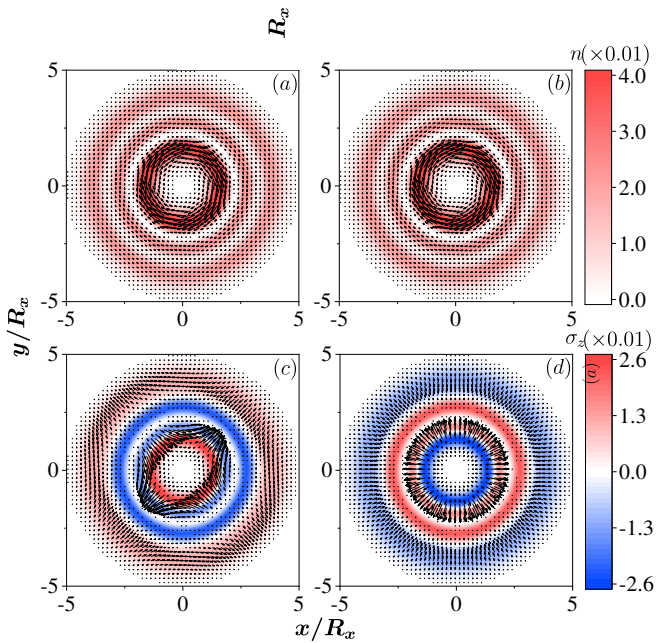}
\caption{(Color online) The density profiles and the spin fields of the $100$th 
eigen states in an isotropic QD ($R_x=R_y=30$ nm) with different SOCs, in the 
absence of external magnetic field. Panels (a) and (c) are for the QD with 
Dresselhaus SOC $\hbar g_2=40$ nm$\cdot$meV, while panels (b) and (d) are for 
the QD with Rashba SOC $\hbar g_1=40$ nm$\cdot$meV. In (a) and (b), the 
color represents the density of the electron and the arrows represent the 
current vector $(j_x(\mathbf{r}), j_y(\mathbf{r}))$. (c) and (d): The color 
stands for $\sigma_z (\mathbf{r})$ and the arrows for the in-plane spin fields 
$(\sigma_x(\mathbf{r}) , \sigma_y(\mathbf{r}))$ with topological charge $-1$ 
and $1$, respectively. All the observable quantities are in units of $1/R_x^2$
hereafter.}
\label{fig1}
\end{figure}

Our study indicates that the densities of all eigenstates have a circular shape with topological 
nontrivial vortex-like spin textures [Fig. \ref{fig1}]. The Rashba SOC 
induces a topological charge $+1$ of the in-plane spin field, while the 
Dresselhaus SOC leads to topological charge $-1$ \cite{luo1,luo2}. Further, the 
current fields of the two cases are also shown in Fig. \ref{fig1}, where the two SOCs 
lead to rotating currents with the same vorticity related to their spin fields. 

When a perpendicular magnetic field is introduced, the electron has a cyclotron 
motion in the magnetic field. The densities of the eigenstates maintain a circular 
structure with rotational symmetry when only one SOC is present in an isotropic QD. 
However, the directions of the current may be changed by the magnetic field in 
different eigenstates.

\subsection{Isotropic QD with combination of different SOCs}

The chaotic dynamics can be driven in the isotropic dot by combining the two 
SOCs arbitrarily and $|g_1| \neq |g_2|$. The classical Hamiltonian is 
\begin{eqnarray}
H^C &=& \frac{\left( p_{x}^{\prime }\right) ^{2}+\left( p_{y}^{\prime }\right)
^{2}}{2m^*}+\frac{1}{2}m^* \omega ^{2}\left( x^{2}+y^{2}\right) \notag \\
&-&\sqrt{\left(
g_{1}+g_{2}\right) ^{2}\left( p_{x}^{\prime }\right) ^{2}+\left(
g_{1}-g_{2}\right) ^{2}\left( p_{y}^{\prime }\right) ^{2}},
\end{eqnarray}
where $p'_x = (p_y - p_x)/\sqrt{2}$ and $p'_y = (p_x + p_y)/\sqrt{2}$. It is obvious 
that the Hamiltonian governs a linear system only when $g_1 = \pm g_2$, since its 
canonical equations are linear. Otherwise, the canonical equations are nonlinear 
and such systems are possible to hold the chaotic dynamics. The isotropically 
confined QD becomes to an anisotropic system due to the arbitrary mixing of the 
two SOCs. This implies that, in the quantum regime, the quantum scars which is 
represented by the electron density localizing along the classical trajectory can 
appear in the excited states. The absence of the magnetic field conserves the 
time reversal symmetry and the quantum scar states appear in pair due to the 
Kramers pair.  

\begin{figure}[htp]
\includegraphics[width=\columnwidth]{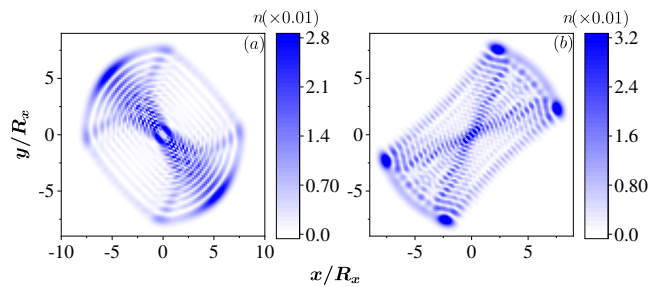}
\caption{(Color online) The quantum scar states in an isotropic QD ($R_x=R_y=30$ 
nm) with mixing of the two SOCs, $\hbar g_1=40$ nm$\cdot$meV and $\hbar g_2
=10$ nm$\cdot$meV. In (a) and (b), the 
color represents the densities of the electron $n(\mathbf{r})$ in the 1028th and 
1247th eigenstates, respectively. }
\label{fig2}
\end{figure}

In Fig. \ref{fig2}, we show two quantum scars in the excited states of an 
isotropic QD with a combination of Rashba and Dresselhaus SOCs, 
$\hbar g_1=40$ nm$\cdot$meV and $\hbar g_2=10$ nm$\cdot$meV. In Fig. 
\ref{fig2}(a), the density of electron is localized to an axe-shape pattern, while 
an `X'-trajectory appears in Fig. \ref{fig2}(b). These patterns are different from the 
array-shaped density profiles of states in a QD without SOC significantly (i.e. the 
densities observed in a 2D quantum oscillator).

\subsection{Quantum Lissajous scar in anisotropic dot induced by a single SOC}

Another quantum scar, which is called quantum Lissajous scar \cite{qls}, can 
emerge in an anisotropic QD where the ratio of the 2D confinements 
$\omega_x/\omega_y$ is a rational number. The two confinement potentials are 
accessible to be manipulated via gates. The original idea to realize the quantum 
Lissajous scars is by the massive random impurities which induce chaos and mix 
different eigenstates of the basis. The scar indicates the classical behavior of an 
anisotropic 2D oscillator, so that the density of the electron of the quantum scar 
state localizes around the Lissajous curve corresponding to the ratio $\omega_x/ \omega_y$.

\begin{figure}[htp]
\includegraphics[scale=0.35]{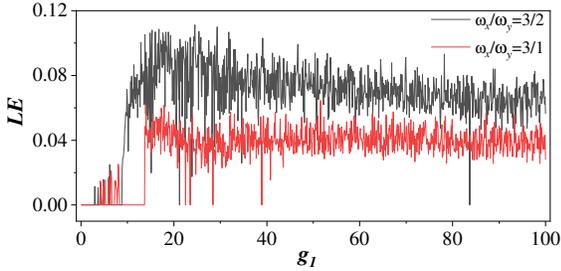}
\caption{(Color online) The largest LEs of the two anisotropic systems with $\omega_x/ \omega_y
=3/1$ and $3/2$. These LEs are calculated in the dimensionless Hamiltonian in Eq. 
(\ref{dimensionless}) with varied $g_1$ and fixed $g_2=0$. The chaos 
of the system is related to $g_1$.}
\label{fig3}
\end{figure}

In an anisotropic QD with Rashba or Dresselhaus SOC, the corresponding classical 
Hamiltonian also leads to chaotic dynamics in the phase space obtained by its 
Hamilton's equation. For simplicity, the dimensionless Hamiltonian with $m^*=1$ is 
\begin{equation}
H^C=  p_x^2 + p_y^2 + \frac{1}{2} 
\omega_x^2 x^2 + \frac{1}{2} \omega_y^2 y^2 - g_{1(2)} \sqrt{p_x^2 + p_y^2},
\label{dimensionless}
\end{equation} 
without the vector potential, i.e. no magnetic field. 
The Lyapunov exponent (LE) is employed to estimate the oscillation modes 
under parameter variation. The largest LE being positive indicates the existence of 
a chaotic state, while the largest LE being negative denotes the system described by 
periodical states only. In Fig. \ref{fig3}, the largest LE \cite{yao}
of the two examples with $\omega_x / \omega_y =3/1,3/2$ demonstrate chaos 
in the system, when $g_2=0$ and $g_1$ is tuned (equivalent to tuning 
energy of the system). Note that for some $g_1$ the system shows no chaos.

\begin{figure}[htp]
\includegraphics[width=\columnwidth]{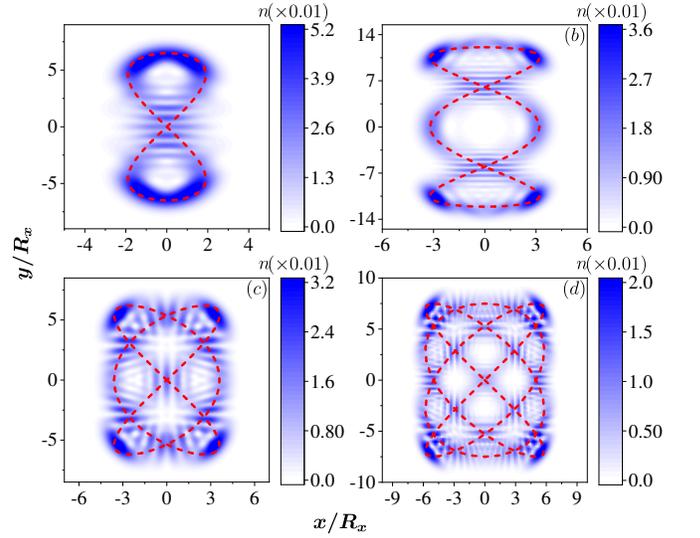}
\caption{(Color online) The quantum scar states in an anisotropic QD 
($R_x=30$ nm) with only the Rashba SOC, $\hbar g_1=40$ nm$\cdot$meV. 
Colors represent for the density of the electron, $n(\mathbf{r})$. (a) The quantum 
scar in the $91$st eigenstate around the Lissajous curve $\sim(\sin 2t, \sin t)$ since 
$R_y=\sqrt{2} R_x$ and $\omega_x / \omega_y =2/1$. (b) For the QD with 
$R_y=\sqrt{3} R_x$ and $\omega_x / \omega_y =3/1$, the quantum scar 
in the $535$th eigenstate around the Lissajous curve $\sim (\sin 3t, \sin (t+\pi/2))$.
(c) For the QD with $R_y=\sqrt{3/2} R_x$ and $\omega_x / \omega_y =3/2$, 
the quantum scar in the $331$st eigenstate around the Lissajous curve 
$\sim(\sin 3t, \sin 2t))$.
(d) For the QD with $R_y=\sqrt{4/3} R_x$ and $\omega_x / \omega_y =4/3$, 
the quantum scar in the $1404$th eigenstate around the Lissajous curve 
$\sim(\sin 4t, \sin 3t))$. The dashed lines are the corresponding Lissajous curves 
drawn for guidance. }
\label{fig4}
\end{figure}

We then demonstrate that the quantum Lissajous scars can be achieved by the 
relativistic correction, i.e., the SOC. In the quantum regime, the emerging quantum scars 
display the trajectory of a particle confined in a classical 2D oscillator. We first discuss the 
scars related to the specific closed Lissajous
curves, $(x,y)\sim \left( \cos \eta_x t, \cos(\eta_y t+\frac{\pi}{2\eta_x}) \right)$, where $\omega_{x,y}=\eta_{x,y}\omega_0$. The open curve obtained 
by shifting the phase will be discussed in the next subsection. The quantum Lissajous scars for $\omega_x/ \omega_y=2/1, 3/1, 3/2, 4/3$ are shown in Figs. \ref{fig4}(a)-(d), respectively. 
Around the cross points in the curves, the interference stripes are clearly visible, 
indicating both the classical and quantum features.  {In addition, the density 
profiles of the first 2000 states in different cases that the confinement ratios and the SOCs are varied} are integrated into a video, which 
can be found in Supplementary Material \cite{video}.

\begin{figure}[htp]
\includegraphics[width=\columnwidth]{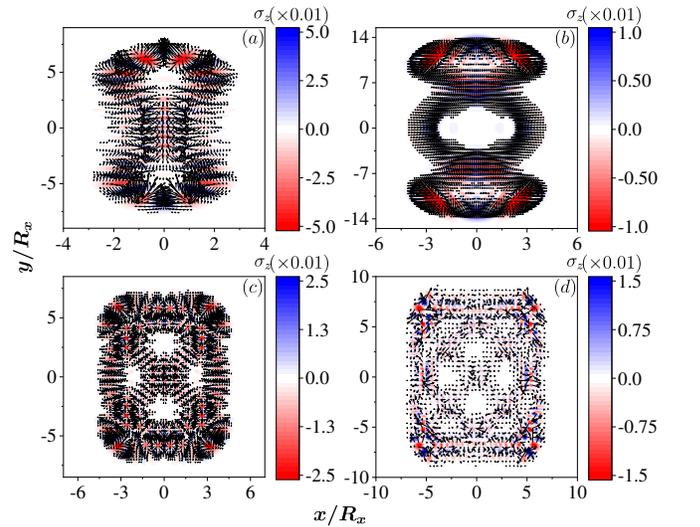}
\caption{(Color online) The spin fields of the quantum scar states in the anisotropic 
QD with the same systematic parameters as those used in Fig. \ref{fig4}. Colors 
represent $\sigma_z (\mathbf{r})$ and the vectors represent the in-plane spin 
fields $(\sigma_x (\mathbf{r}), \sigma_y (\mathbf{r}))$.}
\label{fig5}
\end{figure}

In Figs. \ref{fig5}(a) - (d), we also indicate the associated spin fields of the four quantum Lissajous scars 
[Figs. \ref{fig4}(a) - (d)], respectively. Although the spin textures are 
somehow difficult to calculate analytically given that the perturbation calculations become complex and are not valid with a strong SOC, we can still numerically determine that the 
in-plane spins exhibit nontrivial patterns. There are numerous spin vortices localized and 
attached with the density profile, which are textured by the SOC. 

It is worth noting that some eigenstates do not show any different density profile 
other than the regular dot-array patterns of the 2D quantum oscillator without the SOC. 
It is because in the 
corresponding energy region, the classical dynamics can be regular without 
chaos \cite{SOCchaos}, resulting in the absence of quantum scar states.
The chaotic behavior induced by the SOC differs significantly from that induced 
by random impurities, and so are the quantum scar states. 
Due to the randomness of the impurities sizes and locations, the quantum 
scar states therein can not be controlled or tracked precisely, 
and only the percentage of scar states among all eigenstates can be approximately 
estimated. 

One cannot predict where the quantum scar states induced by impurities are, which makes 
detection of the scar states challenging. In contrast, in an anisotropic QD with SOC, 
the emerging quantum Lissajous scars are not random and can be accurately predicted. 
Each two quantum Lissajous scar states (due to the Kramers pair) appear
quasi-periodically in a few eigenstates. For instance, in the case of 
$\omega_x/\omega_y=3/2$ with $\hbar g_1=40$ nm$\cdot$meV, the quantum scar 
states with density profile similar as those shown in Fig. \ref{fig4}(c) appear repeatedly in 
the ($157$th, $158$nd), ($167$th, $168$th), 
($177$th, $178$th), ($189$th, $190$th),  ($199$th, $200$th) eigenstates, with a period of approximate 10 states between the two pairs of quantum Lissajous scar states.
In higher energies, the Lissajous scar states appear in the 
($303$rd, $304$th), ($317$th, $318$th), ($331$st, $332$nd), ($347$th, $348$th), 
($361$st, $362$nd), ($377$th, $378$th) eigenstates. The separation  
between the two pairs of the quantum Lissajous scar states becomes about 14. 
The period is not fixed and will gradually increase (not monotonically) with 
increase of the energy.

Moreover, the quantum Lissajous scars induced by 
SOC can be found at very low energies, such as the `$8$' shape Lissajous 
trajectory shown in Fig. \ref{fig4}(a), which can be identified even down to the $15$th 
eigenstate. More importantly, the Rashba SOC can be controlled by an 
external gate allowing for the manipulation of the quantum scar states. These 
characteristics  
of the quantum scars induced by SOC imply that SOC, especially the 
tunable Rashba SOC, greatly facilitates the measurement of the quantum scar state.

\begin{figure}[htp]
\includegraphics[width=\columnwidth]{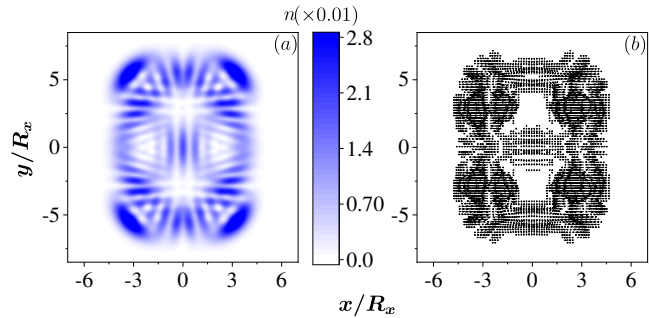}
\caption{(Color online) The quantum scar state in an anisotropic QD ($R_x=30$ 
nm and $R_y=\sqrt{3/2} R_x$) with the Rashba SOC, $\hbar g_1=40.4$ 
nm$\cdot$meV which is a little deviated from that used in Fig. \ref{fig4}(c). 
(a) The quantum Lissajous scar in $331$st eigenstate is the same (both the 
number of the eigenstate and the density profile) as that shown in Fig. \ref{fig4}(c). 
(b) The current field of the $331$st eigenstate.}
\label{fig6}
\end{figure}

Considering the nature of the classical chaotic dynamics being sensitive to initial 
conditions, one might wonder if the quantum scar states are similarly 
sensitive to system parameters. 
If not, then the quantum scar states are more easily detected. We adjust 
the confinement ratio slightly, for instance, $\omega_x / \omega_y=3/2 \rightarrow 
3.01/2$, and observe that the positions of the quantum scar states in all 
the eigenstates remain unchanged, as do their density profiles.
We also examine the effect of adding a weak magnetic field, $B=0.05$ T. 
Although the Kramers' degeneracy is lifted, the quantum scars persist in the same eigenstates as in the absence of 
the magnetic field, with only slight changes in density profiles. Similarly, when the 
SOC strength $g_1$ is slightly tuned, the positions or 
the density profiles of the quantum scar states remain unchanged. As illustrated 
in Fig. \ref{fig6}(a), when the Rashba SOC is increased by one percent compare to 
that in Fig. \ref{fig4}(c), namely $\hbar g_1=40.4$ nm$\cdot$meV, the quantum 
Lissajous scar does not change at all. We note that the current flow direction, displayed in 
Fig. \ref{fig6}(b), does not align with the classical Lissajous trajectory, but is relevant to 
the spin fields shown in Fig. \ref{fig5}(c). This character underscores the fundamental 
distinction between the classical behavior and the quantum mechanism.

The robustness of the quantum scar states against the external perturbations 
relies on the quantum properties of the system rather than its classical behavior. It 
can also be boiled down to the fact that small perturbations do not significantly alter 
the eigen energies of the eigenstates, allowing the corresponding classical 
behavior to remain within the chaos region, thus the scarring is frozen in the discrete-energy 
quantum system. This feature is also helpful for identifying the quantum scar 
states to make the possible measurement convenient. 

Given that the classical Hamiltonians are identical for both 
Rashba and Dresselhaus SOCs, the density profiles of the quantum 
scar states induced by either one of the two SOCs are indistinguishable. Suppose 
that there are two QDs with the same confinement potentials, but one with Rashba SOC and 
the other with Dresselhaus SOC. The coupling strengths in the two QDs are 
identical, $g_1=g_2$. Our numerical studies indicate that the quantum scar states 
appear in the same position in the eigenstates of both cases, exhibiting exactly the 
same density profiles. However, the spin fields of these two states are different, 
providing a signature to distinguish the types of SOC. 

\subsection{Lissajous curves pair scar}

In anisotropic QDs with one SOC, the Lissajous patterns in open curves can also be 
found in scarring states, albeit with much lower probability. However, due to mirror 
symmetry, $x \rightarrow -x$ and $y \rightarrow -y$ without a magnetic field, a single 
open curve of the Lissajous pattern, which has lower symmetry, can not be found in any 
state. Instead, only a pair of Lissajous curves making up this symmetry emerges in 
a scarred state. 

\begin{figure}[htp]
\includegraphics[width=\columnwidth]{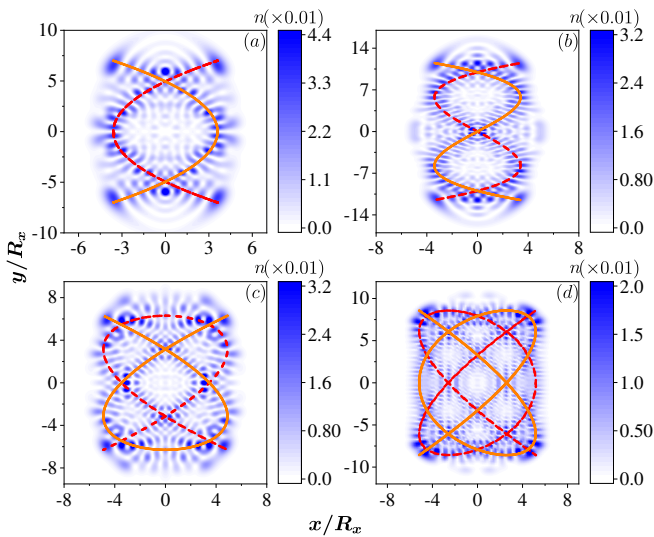}
\caption{(Color online) The Lissajous curves pairs in the quantum scar states in the 
anisotropic QD with the same systematic parameters as those used in Fig. \ref{fig4}. Colors 
represent the density of the electron. The quantum scars in (a) the $417$th eigenstate with 
$\omega_x/\omega_y=2/1$, (b) the $659$th eigenstate with $\omega_x/\omega_y=3/1$, 
(c) the $705$th eigenstate with $\omega_x/\omega_y=3/2$, and (d) the $1571$st eigenstate 
with $\omega_x/\omega_y=4/3$. The lines are the corresponding Lissajous curves 
drawn for guidance.}
\label{fig7}
\end{figure}

In QDs with $\omega_x/\omega_y =2/1, 3/1, 3/2, 4/3$, the pairs of Lissajous curves 
in the quantum scars are illustrated in Figs. \ref{fig7}(a) - (d), where the classical 
orbits $(x,y) \sim (\cos \eta_x t, \cos \eta_y t) + (\cos \eta_x t, \cos (\eta_y t+ \pi/\eta_x))$ 
are identified, respectively.

\subsection{Quantum regular states in anisotropic quantum dots}

\begin{figure}[htp]
\includegraphics[width=\columnwidth]{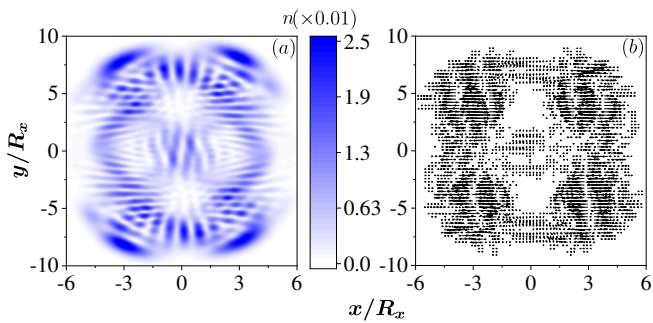}
\caption{(Color online) The quantum scar state in an anisotropic QD with $R_x=30$ 
nm and $R_y=\sqrt{3/2} R_x$. The two SOCs are all present, $\hbar g_1=40$ 
nm$\cdot$meV and $\hbar g_2=10$ nm$\cdot$meV. (a) The quantum Lissajous 
scar in $781$st eigenstate is similar to that in Fig. \ref{fig4}(c), but is a bit twisted. 
(b) The associated current field of this eigenstate.}
\label{fig8}
\end{figure}

Finally, we discuss the effect of combining two SOCs in anisotropic QDs. 
As expected, when $g_1 \neq \pm g_2$, the quantum Lissajous scars appear. In 
Fig. \ref{fig8}, we show that the electron density forms the Lissajous curve in the 
$781$st eigenstate, however, the Lissajous curve is not as regular as the case with 
only one SOC, and is slightly twisted, as does the corresponding current field.  

\begin{figure}[htp]
\includegraphics[width=\columnwidth]{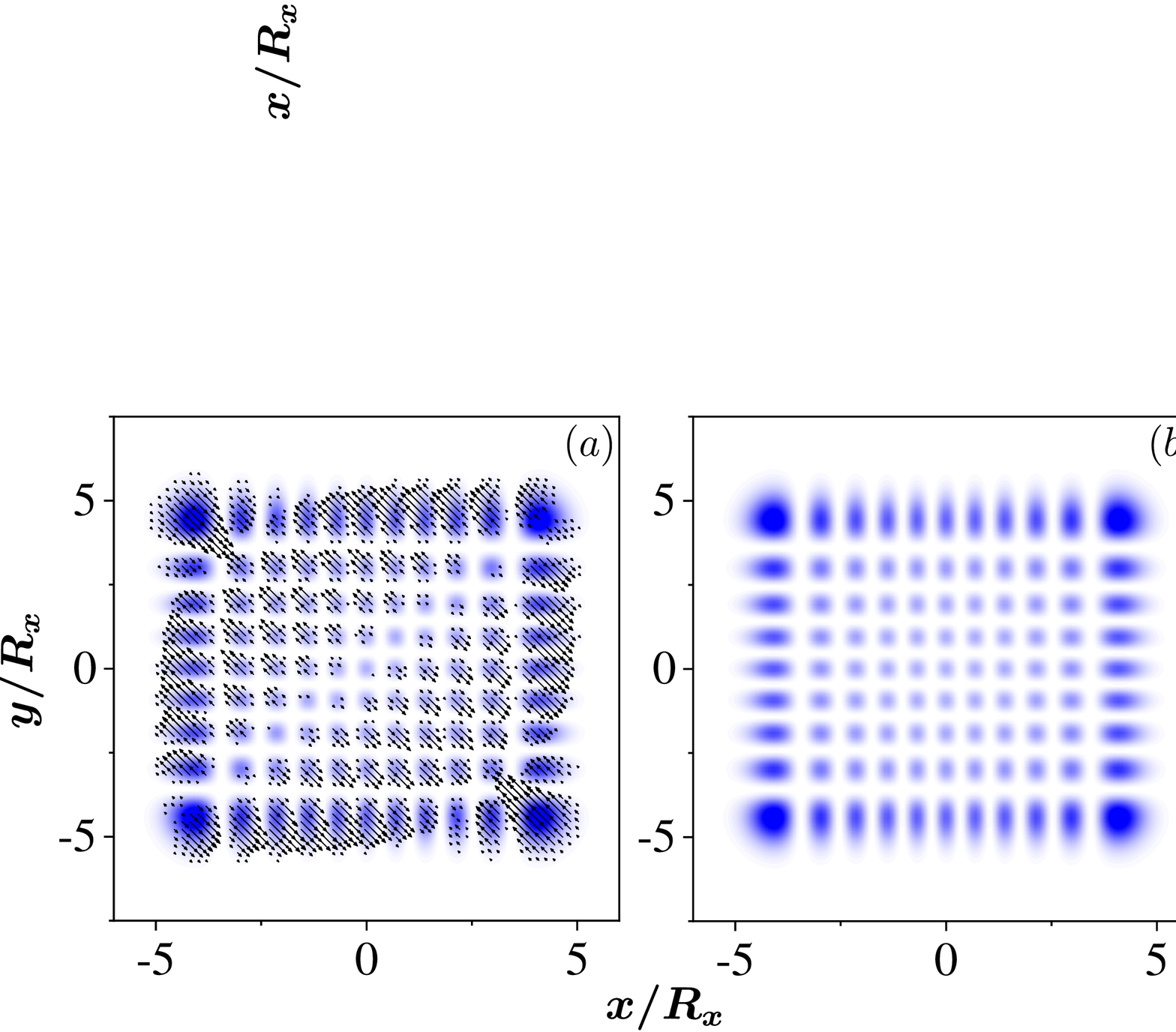}
\caption{(Color online) An example of array-like density in anisotropic QD with 
$R_x=30$ nm and $R_y=\sqrt{3/2} R_x$. The colors represent the density of the 
electron. The $390$th eigenstate is selected for the cases (a) with equal Rashba 
and Dresselhaus SOCs $\hbar g_1=\hbar g_2=10$ nm$\cdot$meV, and (b) 
without SOC. The arrows in (a) represent the in-plane spin field of the state, while 
the in-plane spin field in (b) is zero.}
\label{fig9}
\end{figure}

The special case that $g_1 = g_2$ has the classical correspondence, 
\begin{equation}
H^C = \frac{\left( p_{x}^{\prime }\right) ^{2}+\left( p_{y}^{\prime }\right)
^{2}}{2m^*}+\frac{1}{2}m^* \omega_x^{2}x^{2}++\frac{1}{2}m^* \omega_y^{2} y^{2} 
-2g_1 p'_x,
\end{equation}
which describes a linear system without chaos. A similar Hamiltonian can be 
derived for $g_1=-g_2$. Thus for $g_1 = \pm g_2$ whether the QD is isotropic 
or anisotropic, no classical chaotic dynamics occur and no quantum scar 
appears. Our numerical calculation also confirms that all the density profiles of the 
eigenstates are alike dot-array, which are the same as the densities of the eigenstates 
of the QD without SOC, as shown in Fig. \ref{fig9}. The array-like densities are totally 
induced by the Hermite polynomials in the basis wave functions. The difference of the 
two cases is that the in-plane spin fields are nonzero in the QD with SOCs while the 
spin field is only polarized in the $z$ direction in the QD without SOC.  

When the external magnetic field is weak, the quantum Lissajous scars persist for 
$g_1 \neq \pm g_2$. However, when the magnetic field is increased, the scars are 
overwhelmed by the cyclotron motion. In the case of $g_1=\pm g_2$, due to the 
lack of chaotic dynamics, the densities of all the eigenstates form circles with 
rotational symmetry induced by the magnetic cyclotron motion in an isotropic dot. 
Nevertheless, the densities of the eigenstates remain arrays in an anisotropic QD when 
the magnetic field is weak, but evolve to elliptical shapes with 
increase of the magnetic field. 

\section{Conclusion}
In summary, we have studied the quantum scar states in quantum dots induced by 
relativistic effects, viz. the SOCs. For isotropic quantum dots, only the combination of 
Rashba and Dresselhaus SOCs can induce quantum scars, since the anisotropy 
and chaotic dynamics arise from the interplay between the two SOCs. 
In an anisotropic quantum dot, either one SOC or a combination of the two SOCs 
can lead to quantum Lissajous scar which may consist of one or a pair of 
Lissajous curves. We have to emphasize a special case where $g_1 = \pm g_2$ 
(the two SOCs have the same strength), which corresponds to a linear classical 
system without chaos. Thus, regardless of the confinement of the quantum dot, 
there is no quantum scar appearing in this case. 

The quantum scars induced by SOCs are robust against small perturbations 
of the external conditions, such as small alterations in the confinement ratio 
$\omega_x / \omega_y$, weak magnetic fields, or variation in the strengths of the SOCs. 
The quantum Lissajous scars induced by SOCs emerge quasi-periodically in the 
eigenstates and can manifest at very low energies in particular. It implies that 
tuning the SOC is a stable and controllable way to obtain predictable quantum scars, 
unlike systems where quantum scars induced by a 
bunch of random impurities distribute randomly in the high-energy eigenstates. 
Given that the quantum scars discussed here appear in low-energy states and the 
direct observation of the orbit of the ground state of a quantum dot is already 
realized \cite{observeQD}, our work paves the way to observe the quantum scars 
directly in such nanoscale systems, regardless of the materials, as long as 
the SOC is present. Furthermore, if direct observation is difficult currently, 
other indirect detection methods, such as spin polarization measurements, may 
also be useful due to the robustness of the associated quantum scars and the 
tunable property of the Rashba SOC. Especially, transport signals may be 
utilized to determine the scarring trajectory in quantum dot systems with 
SOCs, and spin-involved transport could prove beneficial for spintronics 
applications.

\section*{Acknowledgement}

W.L. acknowledges Chao Hang and Yu Zhou for helpful discussions. This work is supported by the NSF-China under Grant Nos. 11804396. We are grateful to the High Performance Computing Center of Central South University for partial support of this work.

%%%%%%%%%%%%%%%%%%%%%%%%

\end{document}